\documentclass[aps,prb,twocolumn,showpacs]{revtex4-1}
\usepackage{graphicx}
\usepackage{graphics}
\usepackage{epstopdf}                   

\usepackage[english]{babel}
\usepackage[latin1]{inputenc}           

\usepackage{array}
\usepackage{color}
\usepackage{latexsym}
\usepackage{amsmath,amsxtra,amssymb,amstext,amsfonts}
\usepackage{dsfont}

\begin{document}
\title{An approach to the formalism of the Standard Model
of Particle Physics}

\author{O. E. Casas B.}
\affiliation{Departamento de F\'{\i}sica, Universidad Nacional de Colombia, Bogot\'{a} D. E., Colombia}

\author{A. M. Raba P. and N. Poveda T.}
\affiliation{Escuela de F\'{\i}sica, Universidad Pedag\'{o}gica y Tecnol\'{o}gica de Colombia, Grupo de F\'{\i}sica Te\'{o}rica y Computacional, Tunja, Boyac\'{a}, Colombia}

\begin{abstract}
So far, the Standard Model of particle physics (SM) describes the
phenomenology observed in high energy physics. In the Large Hadron
Collider (LHC) is expected to find the Higgs boson, which is an essential
part of SM; also expects to see new particles or deviations from the
SM, which would be evidence of other truly fundamental theory. Consequently,
a clear understanding of the SM and, in general, quantum field theory
is of great importance for particle physics, however, students face
a formalism and a set of concepts with which they are unfamiliar.
This paper shows how to make an approach to SM to introduce students
to the formalism and some fundamental concepts.
\end{abstract}
\date{\today}
\pacs{01.40.gb, 11.10.Ef, 11.10.Nx}
\maketitle

\section{Introduction}
\label{sec:Introduction}

A problematic situation with the introduction in the study of the
Standard Model of Particle Physics (SM), which constitutes the main
motivation of this article, is the fact that usually the process of
formalizing the model is one of the aspects that cause difficulties
to students. However, it is intended to formalize knowledge, but what
is sought is based on a series of paradigms and rules for creating
a solid foundation on which to work, which can be modified when necessary
\cite{Kuhn}.

\medskip{}

In this context is the SM which describes, with subtle precision and
mathematical elegance, all known particles and interactions, except
gravity. All model predictions have been corroborated, however, the
Large Hadron Collider (LHC) is expected to be a new physics or physics
beyond the SM, in the TeV scale, which can manifest itself in two
ways: through signals involved in the production of new particles
or deviations from SM predictions. Consequently, a clear understanding
of the SM and, in general, quantum field theory is of great importance
for particle physics, this paper shows how to make an approach to
SM to introduce students quickly in the formalism and basic concepts.

\section{Reviewing some basic concepts}
\label{sec:Basic}
\subsection{Fermion-boson interaction}

Everything that exists in nature is made of material particles (called
fermions) which interact, either because they decay or because they
respond to a force due to the presence of another particle. The decay
of a fermion into a different one is explained by the action of a
third call mediator boson. Similarly, the interaction of two particles
through the fields that originate, can be interpreted considering
that the two particles exchange a mediator boson. In quantum mechanics
fermions or bosons can be represented by fields with some intrinsic
properties. The interaction process is an exchange of energy and momentum,
between fields, governed by laws of conservation of intrinsic quantities
(e.g., some kind of charge). 

\subsubsection{Mechanical model}

To explain the interaction between a fermion and a boson, we use a
mechanical model consisting of two coupled mass-spring systems. The Lagrangian of the system is given by:
\[
L=\frac{p_{1}^{2}}{2m_{1}}+\frac{p_{2}^{2}}{2m_{2}}-\frac{1}{2}k.V(x_{1},x_{2}).
\]
The kinetic energy of the first oscillator can represent a free fermion
(e.g., electron), the second oscillator a free boson (e.g., electromagnetic
field) and potential energy, the fermion-boson interaction (e.g.,
electron in electromagnetic field). Note that the interaction is characterized
by a constant (spring constant). Therefore, we will always have three
classes of Lagrangians: the free fermion, the free boson and fermion-boson
interaction. It is clear that the interaction occurs at a predetermined
time, called mean-lifetime.

\subsubsection{Electromagnetic model}

In electromagnetic theory, the Lagrangian for a free particle of mass
$m$ and an free electromagnetic field represented for a potential,
$\overrightarrow{A}$, in vacuum, is written as
\begin{equation}
L_{\textrm{free}}=\frac{p^{2}}{2m}+\frac{1}{8\pi}\int d^{3}x\left(\frac{1}{c^{2}}\left|\frac{\partial\overrightarrow{A}}{\partial t}\right|^{2}+\left|\overrightarrow{\nabla}\times\overrightarrow{A}\right|^{2}\right).\label{eq:Lagrangian-EM-free}
\end{equation}
The introduction of the interaction reduces simply to the
replacement of ordinary momentum by canonical momentum (as is usual
in electrodynamics): $\overrightarrow{P}\rightarrow\overrightarrow{p}+\frac{q}{c}\overrightarrow{A}$.
Substituting we obtain the Lagrangian for a particle interacting:
\[
L_{\textrm{with interaction}}=L_{\textrm{free}}+\frac{q}{2mc}\overrightarrow{A}.\left(2\overrightarrow{p}+\frac{q}{c}\overrightarrow{A}\right).
\]

\subsubsection{Quantum-mechanical model}

In the formalism of relativistic quantum mechanics a free particle
can be represented by a field called spinor,$\Psi=\Psi(x^{\mu})$.
In the continuum, the Lagrangian ($L=L(q,\dot{q})$) is replaced by
a Lagrangian density ($\mathcal{L}$), the generalized coordinates
($q$) by the fields ($\Psi$) and the generalized velocities ($\dot{q}=dq/dt$)
by the field gradient ($\partial_{\mu}\Psi=\partial\Psi/\partial x^{\mu}$):
$\mathcal{L}=\mathcal{L}(\Psi,\partial_{\mu}\Psi)$. For simplicity,
in what follows, we use natural units where $\hbar=1$ and $c=1$.
The Lagrangian density for a free particle and an electromagnetic
field, equivalent to the equation \eqref{eq:Lagrangian-EM-free},
fully consistent with the principles of quantum mechanics and special
relativity ($v\simeq c$), is given by:
\begin{equation}
\mathcal{L}_{\textrm{free}}=\overline{\Psi}(i\gamma^{\mu}\partial_{\mu}-m)\Psi-\frac{1}{4}F_{\mu\nu}F^{\mu\nu},\label{eq:Lagrangian-fermion-foton}
\end{equation}
\noindent where $\overline{\Psi}$ is defined as $\overline{\Psi}=\Psi^{\dagger}\gamma^{0}$
and $F_{\mu\nu}=\partial_{\mu}A_{\nu}-\partial_{\nu}A_{\mu}$. The
first term is called the Dirac Lagrangian. In order to introduce the
interaction we replace the ordinary momentum $p_{\mu}$ by canonical
momentum $P_{\mu}\rightarrow p_{\mu}+qA_{\mu}$, where $q$ is the
coupling constant, which in this case corresponds to the charge. In
language operator the momentum of the particle is given by $p_{\mu}=-i\partial_{\mu}$,
and $P_{\mu}=-iD_{\mu}$, then, the introduction of the interaction
for the relativistic quantum particle interacting
with the field, translates to replace the normal derivative by the covariant derivative in the Lagrangian,
$D_{\mu}\rightarrow\partial_{\mu}-iqA_{\mu}$:
\begin{equation}
\mathcal{\mathcal{L}_{\textrm{\textrm{with interaction}}}}=\mathcal{L}_{\textrm{free}}+q\overline{\Psi}\gamma^{\mu}\Psi A_{\mu},\label{eq:Lagrangian-QED}
\end{equation}
\noindent here, $J^{\mu}=q\overline{\Psi}\gamma^{\mu}\Psi$ is a current
satisfying: $\partial_{\mu}J^{\mu}=0$, indicating that it is a conserved
quantity \cite{Quigg}. 

\begin{figure}[H]
\begin{centering}
\includegraphics[scale=0.6]{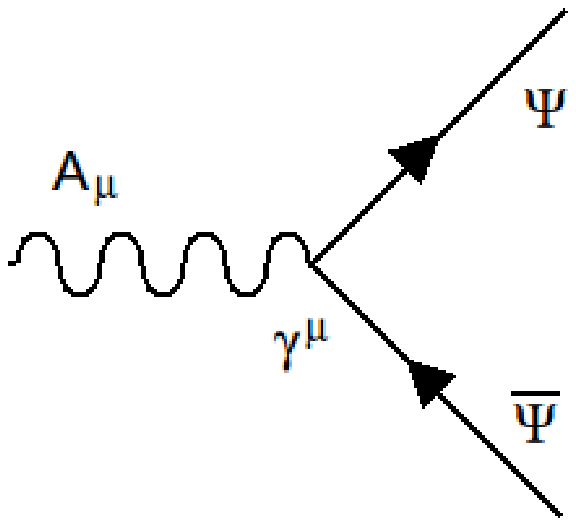}
\par\end{centering}
\caption{Feynman diagram for $\mathcal{\mathcal{L}_{\textrm{i}}}=q\overline{\Psi}\gamma^{\mu}\Psi A_{\mu}$}
\end{figure}

The interaction terms can be represented by Feynman diagrams,
we have six possibilities: creation and annihilation of an electron-positron
pair, or absorption of a photon by an electron or a positron.

\subsection{Quantum operators}

Consider the operator rotation, $R$, where an infinitesimal rotation
is defined by $R(\delta\theta)\left|\Psi(\theta)\right\rangle =\left|\Psi(\theta-\delta\theta)\right\rangle $;
when we expand the right side in a Taylor series around $\theta$,
we get, $R(\delta\theta)\left|\Psi(\theta)\right\rangle =\left(I-iJ\delta\theta\right)\left|\Psi(\theta)\right\rangle $.
where the $J$ operator is called the generator of rotations. Consequently,
a rotation is given by an infinite number of infinitesimal rotations
\cite{Lichtenberg}: 
\begin{equation}
R(\theta)=\underset{N\longrightarrow\infty}{l\acute{\imath}m}\left[I-iJ\left(\frac{\theta}{N}\right)\right]^{N}=e^{-iJ\theta}.\label{eq:Rotation-operator}
\end{equation}
It is not difficult to demonstrate that in order to conserve
the probability the $R$ operator must be unitary, $R^{\dagger}R=RR^{\dagger}=I$;
this is true only if your generator is hermitian $J=J^{\dagger}$.

\begin{table}[H]
\begin{centering}
\begin{tabular}{|c|c|c|c|}
\hline 
Symmetry & Operator & Generator & Conservation law\tabularnewline
\hline
\hline 
Temporary evolution & $e^{-iHt}$ & $H$ & Energy\tabularnewline
\hline 
Translation & $e^{ipx}$ & $p$ & Momentum\tabularnewline
\hline 
Rotation & $e^{-iJ\theta}$ & $J$ & Angular Momentum\tabularnewline
\hline 
4D Rotation & $e^{ip_{\mu}x^{\mu}}$ & $p_{\mu}$ & 4-momentum\tabularnewline
\hline 
Gauge & $e^{-iq\alpha(x)}$ & $q$ & Charge\tabularnewline
\hline
\end{tabular}
\par\end{centering}
\caption{Some quantum operators, symmetry and conservation laws.}
\end{table}

All operators of quantum mechanics can be constructed in the same
way and have the form of a rotation transformation. Noether's theorem
establishes that if an action is invariant under a transformation
group (symmetry), there are one or more quantities (constants of motion)
which are associated with these transformations, in other words, to
each symmetry corresponds a conservation law \cite{Goldstein}. We
see that each operator has a generator, and the generators belong
to a particular Lie group. The commutators of these generators are
linear combinations that define the group algebra, meaning the commutation
of two generators of the group produces a third generator of the same
group; this can be shown using the definition \eqref{eq:Rotation-operator}
\cite{Townsend}.

\subsection{Local phase invariance}

We can change the phase of all the fields that represent the particles
(local phase change), $\Psi\rightarrow\Psi'=e^{-iq\alpha(x)}\Psi,$
note that $\alpha$ is a function that depends on the coordinates
$\alpha=\alpha(x)$ and $q$ is the generator. In this case the Dirac
Lagrangian ceases to be invariant:
\[
\mathcal{L}_{D}=\overline{\Psi}(i\gamma^{\mu}\partial_{\mu}-m)\Psi+q\overline{\Psi}\gamma^{\mu}\Psi\partial_{\mu}\alpha\left(x\right).
\]
We can demand that the Lagrangian is invariant under local phase transformation,
by removing how the additional term that appears. By making the particle
to interact with potential $A_{\mu}$, see equation \eqref{eq:Lagrangian-QED},
the Lagrangian is composed of three terms, the Lagrangian particle
to which is added a term due to the potential and another term due
to local phase change:
\begin{eqnarray*}
\mathcal{L}_{D} & = & \overline{\Psi}(i\gamma^{\mu}D_{\mu}-m)\Psi+q\overline{\Psi}\gamma^{\mu}\Psi\partial_{\mu}\alpha\left(x\right)\\
& = &\overline{\Psi}(i\gamma^{\mu}\partial_{\mu}-m)\Psi+q\overline{\Psi}\gamma^{\mu}\Psi A_{\mu}+q\overline{\Psi}\gamma^{\mu}\Psi\partial_{\mu}\alpha\left(x\right).
\end{eqnarray*}
We can calibrate the potential in such a way that the term of change
of phase is cancelled:
\begin{equation}
A_{\mu}\rightarrow A_{\mu}-q\partial_{\mu}\alpha(x),\label{eq:Potential-gauge}
\end{equation}
\noindent then
\begin{equation}
\mathcal{L}_{D}=\overline{\Psi}(i\gamma^{\mu}D_{\mu}-m)\Psi.\label{eq:Lagrangian-Dirac-gauge}
\end{equation}
We see that this Lagrangian, which represents an interacting particle
with a potential; is invariant under a local phase change, if the
potential is calibrated \eqref{eq:Potential-gauge}. The fields represented
by such calibrated potentials are called gauge fields.

\section{Yang-Mills fields}
\label{sec:Yang-Mills}

Consider that we have $n$ fields represented by potentials, $A_{\mu}^{a}$
$(a=1,2,3,\ldots,n)$, the covariant derivative takes the form: $D_{\mu}=\partial_{\mu}-iq_{a}A_{\mu}^{a}$
and gauge invariance occurs under the local transformation $\Psi\rightarrow\Psi'=e^{-iq_{a}\alpha^{a}(x)}\Psi$.
By means of the operator of change of phase the representation of
the Lie group can be obtained doing $q_{a}=gT_{a}$, where $g$ corresponds
to the coupling constant and $T_{a}$ to the group generators, i.e.,
$\Psi\rightarrow\Psi'=e^{-igT_{a}\alpha^{a}(x)}\Psi$. In this way
we associate a potential $A_{\mu}^{a}$ a group represented by its
generator $T_{a}$. The commutators of the generators of a Lie group
is a linear combination of generators: $\left[T_{a},T_{b}\right]=if_{abc}T^{c}$,
where the coefficients of the linear combination $f_{abc}$, are the
structure constants of the group.\medskip
By analogy with the electromagnetic field tensor, the field tensor
$F_{\mu\nu}^{a}$ is defined as generalized as:
\begin{equation}
iq_{a}F_{\mu\nu}^{a} = \left[D_{\nu},D_{\mu}\right].\label{eq:Tensor-Yang-Mills}
\end{equation}
Substituting the covariant derivative and commutator of the generators
of the group is obtained:
\begin{equation}
F_{\mu\nu}^{a}=\partial_{\mu}A_{\nu}^{a}-\partial_{\nu}A_{\mu}^{a}+gf_{bc}^{a}A_{\mu}^{b}A_{\nu}^{c},\label{eq:Tensor de campo Yang-Mills}
\end{equation}
\noindent we see that the first two terms of this expression correspond
to a tensor analogous to the electromagnetic field and the last term
is the fact that generators are not abelian group; this allows to
create no-abelians gauge theories or Yang-Mills theories \cite{Yang-Mills}.
With the gauge field tensor is constructed from Yang-Mills Lagrangian
for the free field:
\begin{equation}
\mathcal{L}=-\frac{1}{4}F_{\mu\nu}^{a}F_{a}^{\mu\nu},\label{eq:Lagrangiano campo libre Yang-Mills}
\end{equation}
\noindent note its similarity to the free electromagnetic field Lagrangian
\eqref{eq:Lagrangian-fermion-foton}.

\section{The Standard Model}
\label{sec:SM}

\subsection{Strong Interaction}

To build the model for the strong interaction we propose a Lagrangian
invariant under local gauge transformations of the $SU(3)_{C}$ group,
the subscript $C$ indicates that the transformations only act on
particles with color charge. The strong interaction is represented
by means eight gauge fields for gluons $G_{\mu}^{a}$ $(a=1,2,\ldots,8)$
and each of these fields is associated generators $T_{a}=\lambda_{a}/2$,
being $\lambda_{a}$ the Gell-Mann matrices: $\Psi\rightarrow\Psi'=e^{-ig_{s}T_{a}\alpha^{a}(x)}\Psi$.
The coupling constants of the fields with gluons are $q_{a}=g_{s}T_{a}$.
Lie algebra, is obtained with the following rule of commutation: $\left[T_{a},T_{b}\right]=if_{ab}^{c}T_{c}$,
where $f_{ab}^{c}$ it is the structure constant of the $SU(3)$ group.
The covariant derivative is given by:
\[
D_{\mu}=I\partial_{\mu}-ig_{s}\frac{\lambda_{a}}{2}G_{\mu}^{a}.
\]
Applying the definition \eqref{eq:Tensor-Yang-Mills}\eqref{eq:Tensor de campo Yang-Mills}
obtain the tensor field:
\[
G_{\mu\nu}^{a}=\partial_{\mu}G_{\nu}^{a}-\partial_{\nu}G_{\mu}^{a}+g_{s}f_{bc}^{a}G_{\mu}^{b}G_{\nu}^{c},
\]
\noindent and for the free field Lagrangian \eqref{eq:Lagrangiano campo libre Yang-Mills}
is 
\[
\mathcal{L}=-\frac{1}{4}G_{\mu\nu}^{a}G_{a}^{\mu\nu}.
\]

\subsection{Electroweak Interaction}

A very interesting aspect of SM is that the electromagnetic interactions
and weak interactions are combined into a single, so called, electroweak
interaction. For electroweak interactions we propose a Lagrangian
invariant under local gauge transformations of the $SU(2)_{L}\otimes U(1)_{Y}$
group, the subscript $L$ means that these transformations only act
on components of the left chiral fermions and $Y$ is the hypercharge
quantum number. Electroweak interactions are represented by four gauge
fields $W_{\mu}^{i}$ $(i=1,2,3)$ and $B_{\mu}$. Each of these fields
are associated generators $\tau_{i}=\sigma_{i}/2$ of the weak isospin
group $SU(2)_{L}$ and $y=IY/2$, corresponding to the $U(1)_{Y}$
group, respectively. Here $\sigma_{i}$ are the Pauli matrices, $I$
the $2\times2$ unit matrix and $Y$ hypercharge quantum number. The
coupling constants of the flavor change with the electroweak field
are $q_{i}=g\tau_{i}$ and $q=g'y$. The group algebra is given by
the following commutation rules: $\left[\tau_{i},\tau_{j}\right]=i\epsilon_{ij}^{k}\tau_{k}$
and $\left[\tau_{i},y\right]=0$, where $\epsilon_{ij}^{k}$ is the
constant of structure of the $SU(2)$ group. The covariant derivative
is given by:
\begin{equation}
D_{\mu}=I\partial_{\mu}-ig\frac{\sigma_{i}}{2}W_{\mu}^{i}-ig'I\frac{Y}{2}B_{\mu}.\label{eq:Derivada gauge electrodebil}
\end{equation}
Applying again the definition \eqref{eq:Tensor-Yang-Mills}\eqref{eq:Tensor de campo Yang-Mills}
one obtains the tensor field:
\begin{eqnarray*}
W_{\mu\nu}^{i} &=& \partial_{\mu}W_{\nu}^{i}-\partial_{\nu}W_{\mu}^{i}+gf_{jk}^{i}W_{\mu}^{j}W_{\nu}^{k},\\
B_{\mu\nu} &=& \partial_{\mu}B_{\nu}-\partial_{\nu}B_{\mu},
\end{eqnarray*}
\noindent note that the gauge fields $W_{\mu}^{i}$ are non-abelian
Yang-Mills fields and gauge field $B_{\mu}$ is a abelian field, because
the structure constant of the group is zero. For the free field \eqref{eq:Lagrangiano campo libre Yang-Mills}
is 
\[
\mathcal{L}=-\frac{1}{4}W_{\mu\nu}^{i}W_{i}^{\mu\nu}-\frac{1}{4}B_{\mu\nu}B^{\mu\nu}.\]
Performing the matrix operations indicated in the covariant derivative
\eqref{eq:Derivada gauge electrodebil} and by means the linear combination
of the first two components of the gauge field, $W_{\mu}^{1}$ y $W_{\mu}^{2}$,
one can build charged intermediate vector bosons, $W_{\mu}^{+}$
y $W_{\mu}^{-}$:
\begin{equation}
W_{\mu}^{\pm}=\frac{1}{\sqrt{2}}\left(W_{\mu}^{1}\mp iW_{\mu}^{2}\right),\label{eq:Campos cargados W}
\end{equation}
\noindent and with the mixture of the third component of the gauge
field, $W_{\mu}^{3}$ with the gauge field $B_{\mu}$, the neutral
vector boson is obtained $Z_{\mu}^{o}$ and the boson of the electromagnetic
field, $A_{\mu}^{o}$ (photon) (superscript $o$ indicates that it
is a field with neutral electric charge):
\begin{equation}
\left(\begin{array}{c}
Z_{\mu}^{o}\\
A_{\mu}^{o}\end{array}\right)=\left(\begin{array}{cc}
\cos\theta_{w} & -\sin\theta_{w}\\
\sin\theta_{w} & \cos\theta_{w}\end{array}\right)\left(\begin{array}{c}
W_{\mu}^{3}\\
B_{\mu}\end{array}\right),\label{eq:Campos neutros Z y A}
\end{equation}
\noindent where the mixing angle, $\theta_{w}$, called Weinberg angle;
which is defined by the ratio of the electroweak coupling constants,
$\tan\theta_{w}=g'/g$ and the value of the electron charge, $e=g\sin\theta_{w}=g'\cos\theta_{w}$,
which represents the unification of the weak field with the electromagnetic
\cite{Quigg}. The hypercharge $Y$ is related to the electric charge
$Q$ by the Gell-Mann Nishijima relation \cite{Gell-Mann}: $Q=\tau_{3}+\frac{1}{2}IY$,
here $\tau_{3}=\sigma_{3}/2$ it is third generator of the $SU(2)$
group and $Y$ hypercharge quantum number. Substituting these definitions,
\eqref{eq:Campos cargados W} and \eqref{eq:Campos neutros Z y A}
in the covariant derivative \eqref{eq:Derivada gauge electrodebil},
we obtain:
\begin{equation}
\begin{array} {c c c}
D_{\mu} &=& I\partial_{\mu}-ieQA_{\mu}^{o}-i\frac{g}{\sqrt{2}}\sigma^{+}W_{\mu}^{+}-i\frac{g}{\sqrt{2}}\sigma^{-}W_{\mu}^{-}\\
 & & -i\frac{g}{\cos\theta_{w}}\left(\tau_{3}-Q\sin^{2}\theta_{w}\right)Z_{\mu}^{o},
\end{array}
\label{eq:Derivada covariante electrodebil} 
\end{equation}
\noindent where $Q=\frac{1}{2}\left(\begin{array}{cc}
Y+1 & 0\\
0 & Y-1\end{array}\right)$.

\subsection{Fermions}

Fields representing fermions can be decomposed into a left and right
chiral components: $\Psi=\Psi_{L}+\Psi_{R}$, using chiral projectors
$P_{L}=\frac{1}{2}(1-\gamma^{5})$ and $P_{R}=\frac{1}{2}(1+\gamma^{5})$,
where $\Psi_{L}=P_{L}\Psi$ and $\Psi_{R}=P_{R}\Psi$.\medskip

Fermions, quarks and leptons, are classified into three groups called
generations. A generation is formed by quarks and an leptons with
different charges, and generations are ordered from the lightest to
the heaviest. The left component of the leptons is given by a doublet
$\Psi_{Li}=\left(\begin{array}{c}
\nu_{i}\\
e_{i}\end{array}\right)_{L}$, where $\nu_{i}\equiv\nu_{e},\nu_{\mu},\nu_{\tau}$ and $e_{i}\equiv e,\mu,\tau$;
for quarks $\Upsilon_{Li}=\left(\begin{array}{c} u_{i}\\
d'_{i}\end{array}\right)_{L}$, where $u_{i}\equiv u,c,t$; $d'_{i}=V_{ij}d_{j}$ with $V_{ij}$
an element of the Cabibbo-Kobayashi-Maskawa (CKM) matrix and $d_{j}=d,s,b$.
The CKM mixing matrix is due to the fact that weak interaction does
not act equally on quarks but is divided between them, however, the
theoretical origin of this mixture has not been determined yet. The
right component of the leptons is given by a singlet: $\Psi_{Ri}=e_{iR}$,
because in the SM neutrinos are massless, they do not have a right chiral
component for them; while for the quarks $\Upsilon_{Ri}\equiv u_{iR},d_{iR}$.\medskip

To describe the leptons and quarks interacting with the weak and electromagnetic
fields, we use the Lagrangian \eqref{eq:Lagrangian-leptonic}. As
the left chiral components ($\Psi_{L}$) and right ($\Psi_{R}$) have
different local phase transformations, the generator of the left component
belongs to the $SU(2)$ group, while the right ones only to the $U(1)$
group, the term mass $m\overline{\Psi}\Psi=m\left(\overline{\Psi}_{R}\Psi_{L}+\overline{\Psi}_{L}\Psi_{R}\right)$
is not gauge invariant, thus is absent in the Lagrangian:
\begin{equation}
\mathcal{L}_{l}=\overline{\Psi_{Li}}i\gamma^{\mu}D_{L\mu}\Psi_{Li}+\overline{\Psi_{Ri}}i\gamma^{\mu}D_{R\mu}\Psi_{Ri},\label{eq:Lagrangian-leptonic}
\end{equation}
\noindent where covariant derivatives are given by $D_{L\mu}=D_{\mu}$
\eqref{eq:Derivada covariante electrodebil} and $D_{R\mu}=\partial_{\mu}-ieQA_{\mu}^{o}+i\frac{g}{\cos\theta_{w}}Q\sin^{2}\theta_{w}Z_{\mu}^{o}$.
Consequently, the Lagrangian describes only massless leptons interacting
with the weak and electromagnetic fields. Performing operations gives:
\begin{eqnarray*}
\mathcal{L}_{l} & = & i\overline{\nu_{i}}\gamma^{\mu}\partial_{\mu}P_{L}\nu_{i}+i\overline{e_{i}}\gamma^{\mu}\partial_{\mu}e_{i}\\
 &  & -e\overline{e_{i}}\gamma^{\mu}e_{i}A_{\mu}^{o}+\frac{g}{2\cos\theta_{w}}\overline{\nu_{i}}\gamma^{\mu}P_{L}\nu_{i}Z_{\mu}^{o}\\
 &  & +\frac{g}{\sqrt{2}}\overline{\nu_{i}}\gamma^{\mu}P_{L}e_{i}W_{\mu}^{+}+\frac{g}{\sqrt{2}}\overline{e_{i}}\gamma^{\mu}P_{L}\nu_{i}W_{\mu}^{-}\\
 &  & +\frac{g}{2\cos\theta_{w}}\overline{e_{i}}\gamma^{\mu}\left(R_{e}P_{R}+L_{e}P_{L}\right)e_{i}Z_{\mu}^{o},
\end{eqnarray*}
\noindent where $R_{e}=2\sin^{2}\theta_{w}$ and $L_{e}=2\sin^{2}\theta_{w}-1$.
Similarly, the Lagrangian of the quark sector is given by:
\begin{eqnarray*}
\mathcal{L}_{q} & = & i\overline{u_{i}}\gamma^{\mu}\partial_{\mu}u_{i}+i\overline{d_{j}}\gamma^{\mu}\partial_{\mu}d_{j}\\
 &  & +\frac{2e}{3}\overline{u_{i}}\gamma^{\mu}u_{i}A_{\mu}^{o}-\frac{e}{3}\overline{d_{j}}\gamma^{\mu}d_{j}A_{\mu}^{o}\\
 &  & +\frac{g}{\sqrt{2}}V_{ij}\overline{u_{i}}\gamma^{\mu}P_{L}d_{j}W_{\mu}^{+}+\frac{g}{\sqrt{2}}V_{ij}^{\ast}\overline{d_{j}}\gamma^{\mu}P_{L}u_{i}W_{\mu}^{-}\\
 &  & +\frac{g}{\cos\theta_{w}}\overline{u_{i}}\gamma^{\mu}\left(\frac{1}{2}P_{L}-\frac{2}{3}\sin^{2}\theta_{w}\right)u_{i}Z_{\mu}^{o}\\
 &  & +\frac{g}{\cos\theta_{w}}\overline{d_{j}}\gamma^{\mu}\left(-\frac{1}{2}P_{L}+\frac{1}{3}\sin^{2}\theta_{w}\right)d_{j}Z_{\mu}^{o}.
\end{eqnarray*}

\subsection{Terms of mass}

To generate the mass terms, we use the Higgs mechanism, which starts from the introduction of a doublet complex scalar fields $\Phi=\left(\begin{array}{c}
\phi^{+}\\
\phi^{o}\end{array}\right)$. The real part of the neutral field can be divided into two components,
$\phi^{o}=\left(h+v+i\eta\right)/\sqrt{2}$, so that when the expected
value in vacuum is nonzero: $\left\langle \Phi\right\rangle _{o}=\frac{1}{\sqrt{2}}\left(\begin{array}{c}
0\\
v\end{array}\right)$, there is a spontaneous breaking of symmetry (RES) according to the
scheme $SU(2)_{L}\otimes U(1)_{Y}\rightarrow U(1)_{Q}$ where $Q$
is the electromagnetic charge \cite{Griffiths}. To generate the mass
of the bosons, we define the kinetic Lagrangian:
\[
\mathcal{L}_{b}=\left(D_{\mu}\Phi\right)^{\dagger}D^{\mu}\Phi,
\]
\noindent here $D_{\mu}$ is given by (\ref{eq:Derivada covariante electrodebil}),
for the ground state ($\Phi\rightarrow\left\langle \Phi\right\rangle _{o}$),
the Lagrangian becomes: $\mathcal{L}_{b}=M_{W}^{2}W_{\mu}^{+}W^{-\mu}+\frac{1}{2}M_{Z}^{2}Z_{\mu}Z^{\mu}$
where, $M_{w}=gv/2$, is the W boson mass; $M_{Z}=gv/(2\cos\theta_{w})$,
Z boson mass and the $A_{\mu}$ boson remains massless. This makes
the weak interaction a short range and long range electromagnetic
interaction.

\medskip{}

The leptons and quarks acquire mass through interaction between fermion
fields and the Higgs field which is called the Yukawa Lagrangian:

\[
-\mathcal{L}_{Y}=\overline{\Psi_{Li}}\lambda_{\mathrm{ij}}\Phi e_{jR}+\overline{\Upsilon_{Li}}\lambda_{\mathrm{ij}}\widetilde{\Phi}u_{jR}+\overline{\Upsilon_{Li}}\lambda_{\mathrm{ij}}\Phi d_{jR}+\mbox{h.c.},
\]

\noindent where $\widetilde{\Phi}=i\tau_{2}\Phi^{*}$; mass terms
are defined as $m_{ij}=\lambda_{ij}v/\sqrt{2}$. For the ground state
($\Phi\rightarrow\left\langle \Phi\right\rangle _{o}$), the Lagrangian
becomes:

\begin{eqnarray*}\mathcal{L}_{Y} & = & m_{e_{i}}\overline{e_{i}}e_{i}+m_{u_{i}}\overline{u_{i}}u_{i}+m_{d_{i}}\overline{d_{i}}d_{i}\\
 &  & +\frac{g}{2M_{w}}\left(m_{e_{i}}\overline{e_{i}}e_{i}+m_{u_{i}}\overline{u_{i}}u_{i}+m_{d_{i}}\overline{d_{i}}d_{i}\right)h^{o}\\
 &  & +\frac{ig}{2M_{w}}\left(m_{e_{i}}\overline{e_{i}}\gamma^{5}e_{i}-m_{u_{i}}\overline{u_{i}}\gamma^{5}u_{i}+m_{d_{i}}\overline{d_{i}}\gamma^{5}d_{i}\right)\eta^{o}\\
 &  & +\frac{gm_{e_{i}}}{\sqrt{2}M_{w}}\overline{\nu_{i}}P_{R}e_{i}\phi^{+}+\frac{gm_{e_{i}}}{\sqrt{2}M_{w}}\overline{e_{i}}P_{L}\nu_{i}\phi^{-}\\
 &  & +\frac{g}{\sqrt{2}M_{w}}\overline{u_{i}}\left[m_{d_{j}}P_{R}-m_{u_{i}}P_{L}\right]V_{ij}d_{j}\phi^{+}\\
 &  & +\frac{g}{\sqrt{2}M_{w}}\overline{d_{i}}\left[m_{d_{j}}P_{L}-m_{u_{i}}P_{R}\right]V_{ij}^{\ast}u_{j}\phi^{-}.
\end{eqnarray*}

\section{\textbf{Conclusion}}
\label{sec:Discussion}

In this paper, using a small number of concepts, we have obtained the main Standard Model Lagrangians. For brevity, we have omitted some issues such as the spontaneous breaking of symmetry, the Higgs mechanism, the Higgs potential, the Faddeev-Popov Lagrangian and others. However, students can find and investigate some supplementary texts.

\section*{Acknowledgments}

Direction of Investigations (DIN) of the Universidad Pedag\'{o}gica y
Tecnol\'{o}gica de Colombia (UPTC) and the Young Investigator program
of COLCIENCIAS.


\begin{thebibliography}{}

\bibitem{Kuhn}T. S. Kuhn, The Structure of Scientific Revolutions (Univ of Chicago Pr, 1996) 3 ed. 

\bibitem{Quigg}C. Quigg, Gauge theories of the strong, weak, and electromagnetic interactions (Addison-Wesley Publishing Company, NY, 1983).  

\bibitem{Lichtenberg}D. B. Lichtenberg, Unitary symmetry and elementary particles (Academic press, NY, 1970) p.35.

\bibitem{Goldstein}H. Goldstein, Classical Mechanics (Addison-Wesley Company, MA, 1987).

\bibitem{Townsend}J. Townsend S., A modern approach to quantum mechanics (Mc-Graw Hill, NY, 1992) p.64-69.

\bibitem{Yang-Mills}C. N. Yang and R. L. Mills, 1954, Phys. Rev. \textbf{96},
191.

\bibitem{Gell-Mann}M. Gell-Mann, 1956, Phys. Rev. \textbf{92}, 833; K. Nishijima
and T. Nakano, 1953, Prog. Theor. Phys. \textbf{10}, 581. 

\bibitem{Griffiths}D. J. Griffiths, \textit{Introduction to elementary particles} (Wiley, NY, 1987) p.360-368.

\end{thebibliography}
\end{document}